# Effect of laser surface hardening on a low carbon steel substrate wear and hardness

**Matheus Rodrigues Furlani[1], Sheila Medeiros de Carvalho[1], Milton Sergio Fernandes de Lima[2]***

1 Mechanical Engineering Department, Federal University of Espirito Santo, Av. Fernando Ferrari 51, Vitoria - ES, 29075-910 Brazil

2 Photonics Division, Institute for Advanced Studies, Trevo Amarante 1, Sao Jose dos Campos - SP, 12228-001 Brazil

* Corresponding author: miltonmsfl@fab.mil.br

**Abstract**

Laser surface hardening (LSH) is an effective process to produce hard and wear resistant steel surfaces. This work produced hardened surfaces using a fiber laser in a bare (B. uncoated) and carbon-coated (C) conditions. The obtained hardness and microstructure were compatible to martensite (B) and cementite (C) major constituents. A low coefficient of friction was obtained using C-condition. The treated surface also presented little damage during a reciprocating wear against a hard metal sphere.

**Introduction**

Laser surface hardening (LSH) is an efficient method to produce hard and wear resistant surfaces in steels with a very small final deformation, and then suitable to near net shape components [1]. LSH could be fully automatized and integrated to the workflow of many transformation industries, granting speed and reliability to the processed components. In this method a laser beam rapidly heat a given volume above the austenite transformation temperature (A3) and then let it cool down below martensite start temperature by a self-quenching mechanism [2]. The current study intends to contribute to LSH of a low carbon steel AISI 1020 using a fiber laser.

It is already know that the volume of the substrate is important for self-quenching mechanism. So and Ki [3] modeled the heat transport phenomena in AISI 1020 low carbon steel and show that a self-quenchable thickness must be equal or higher than 10 mm. The samples to be analyzed in the present study has 3 mm thickness and then collateral heat effects may appear. The same authors [4] also published some simulations about the case depth as a function of laser parameters for a high power diode laser. Unfortunately, this parameterization do not apply to the current fiber laser surface hardening.

The same laser source and optical setup of the present work was studied by Goia and Lima [5] for the AISI D6 steel. According to the authors, a relatively low laser power (e.g. 150W) and low scanning speed (e.g. 10 mm/s) produced a melt-free hardened surface of approximately 200 µm when the laser beam was defocused to a 2 mm spot diameter. These values will guide further developments in this work. The aforementioned study, however, does not establish the conditions for decarburization due to exposure to the laser beam during hardening. This factor is important for practical reasons as some hardening operations in industry do not require gas shielding, or a gas flux is unpractical and costly [6].

Maharjan et al. [7] reported the effect of the surrounding atmosphere in laser hardening of SAE 1020 substrate. The authors reported the effect of the decarburization on the maximum hardness attained at the material' surface indicating propane as a convenient gas shielding method. However, this gas is highly flammable and could cause healthy issues. This work aimed to an alternative route of pre-coating the exposed surface with colloidal graphite in order to reduce decarburization.

In addition to the different perspectives of the present study in relation to the literature, the reciprocal wear will be studied for AISI 1020 steel surfaces after LSH. The friction wear behavior on low carbon steel samples is not very interesting in practice due to its low hardenability. However, LSH can provide a superior surface from a functional point of view, while keeping the core of the piece unchanged. Therefore, this treatment gives more value to a relatively inexpensive type of steel.

The aim of the present study is determine the wear of AISI 1020 steel after laser treatment using dry reciprocating tribology tests.

**Material and methodology**

A commercial 12.5 mm width square rod of AISI 1020 steel was segmented in 3 mm thick samples, ground and grinded in order to obtain parallel and reproductive surfaces for LSH. The original rod was supplied in an as annealed state with chemical composition given by the standard: carbon 0.17-0.23%, manganese 0.30-0.60% and phosphorus plus sulfur below 0.05% (in weight). The base material microstructure is composed by ferritic grains with average Vickers hardness of 115 HV.

Laser surface hardening was carried out using a ytterbium:glass fiber laser (IPG, YLR 2000) without gas shielding to simulate industrial operations. The sample moving was accomplished by a CNC three axes table. The laser beam with a $M^2$ quality of 9 was defocused by 12.2 mm to obtain a Gaussian diameter (1/e2) of 2 mm on the sample surface. The process parameters were obtained from a previous work, i.e. ref. 5. The laser power was fixed at 150 W and the laser beam speed were 10 mm/s. In order to treat the entire surface, each laser track was displaced by 0.5 mm and, therefore, each subsequent run covered 75% of the previous track. The surfaces were coated using a colloidal graphite spray to avoid decarburization, but some surface were retained in an uncoated condition for comparative purposes.

The samples cross-section were analyzed using standard metallography using diamond paste polishing down to 1 µm. The microstructure revelation was done using Nital 2% solution (2% nitric acid in ethanol). The optical microscope was a Zeiss Axio Imager Z2m, which had been used also to analyses the worn surfaces.

A Rigaku Co Ultima IV X-ray diffractometer was used for analyze the phases on the treated surfaces. A copper anode tube produced a Cuk$\alpha$ radiation with a wavelength of 1.54051 Å and the optical configuration was Bragg-Bretano-type. The crystallographic phase data were obtained using the Crystallographic Open Database [8].

The hardness analysis of the hardened surface and sample cross-sections was obtained by microindentation carried out by Vickers hardness test using a TM-800 Futuretech Microdurometer. The applied load was 100 gf ($HV_{0.1}$) and the dwell time was 10 seconds.

Reciprocating sliding wear tests were carried out using a TRB3 equipment (Anto Paar) using a 3 mm diameter hard metal ball and 2 N normal load under ambient temperature. The scratches presented a full amplitude of 4 mm, with a maximum linear speed of 3 cm/s and frequency of 2.39 Hz. The tests stopped after the ball travels 70 m, or 3500 seconds. The dynamic coefficient of friction was recorded as a function of time and the resulting wear was evaluated in a light optical microscope. Each treated surface was tested three times.

## Results and discussion

### Microstructural analyses

The experiments were carried out under two different conditions keeping constant the laser power (150 W), scanning speed (10 mm/s), focusing distance (for a 2 mm spot diameter) and lateral step of 0.5 mm (75% superposition of the tracks). The conditions are bare (B, i.e. uncoated) and coated (C, i.e. with a graphite layer). For cleanness, these conditions will be called B and C thereafter and the unique difference between B and C is the carbon layer in the latter. As can be seen in Figure 1a-b the morphology of the cross sectional area in both B and C are similar. The hardened region (HR) attained a maximum of 100 µm and 88 µm for the conditions B and C, respectively. These values are comparable to previous case depths for the same laser facility [5]. These values also indicate a contradiction in the sense that the presence of a carbon layer would increase the absorptivity [9] and, therefore, the case depth (HR) should theoretically be deeper with the use of carbon. The same phenomena is noticed for the maximum depth of the partially transformed region (PTR), with the B-condition attaining approximately 300 µm compared to 200 µm of C-condition PTR. A possible explanation is that a portion of the laser beam energy was consumed by phase transformation in a C-rich melt. This is particularly true for the reaction $3Fe + C \rightarrow Fe_3C$ (cementite reaction) where the enthalpy is 15 kJ/mol [10].

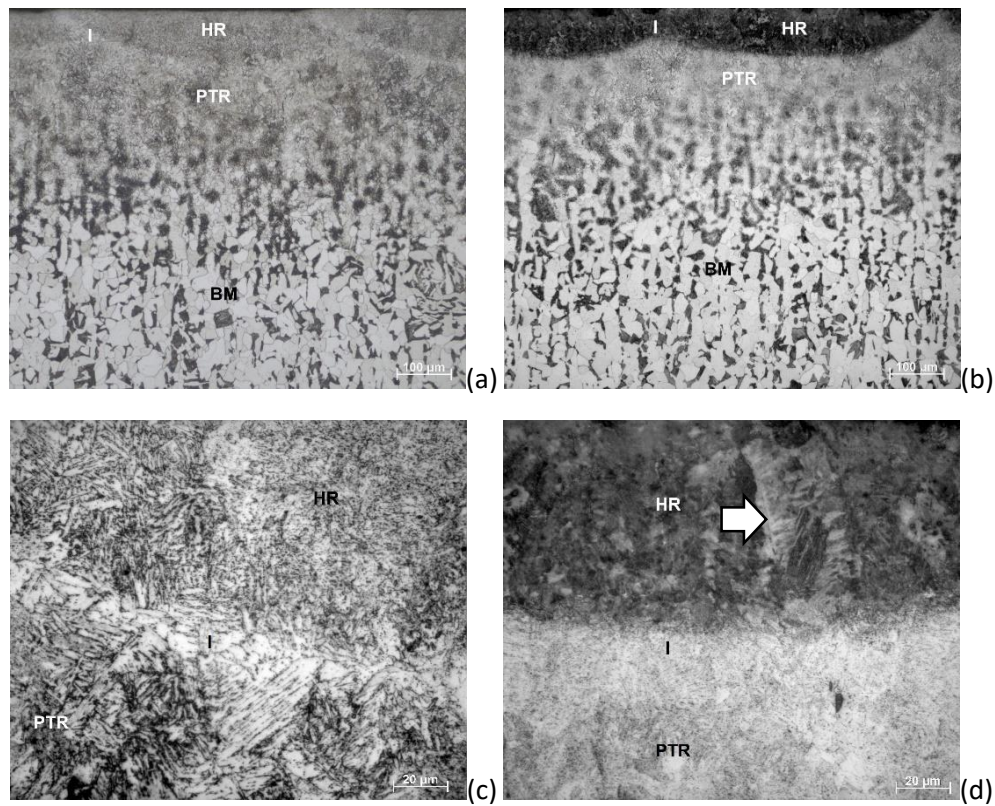

Figure 1. Light optical images of the cross sections of the samples near to the treated surface. (a) low-magnification image of the B-condition; (b) low-magnification image of the C-condition; (c) high-magnification image of the B-condition; (d) high-magnification image of the C-condition. Nital Etching.

Figure 1c-d presents the same regions of Figure 1c-d, but with more magnification. As can be seen in Figure 1c, the microstructure of the bare condition HR and PTR is marked by martensite transformation and the interface (I) presented tempered martensite due to the partial overlap between the tracks. Figure 1d presents a zoom of the region near to the treated surface for condition C. Both PTR and I are composed of martensite, however the HR clearly differ from B-condition. A faceted micro-constituent linked to the cementite phase occurred (marked by an arrow in Figure 1d) corroborating the assumption of carbide phase growth in HR when graphite was applied to the surface. Looking at the interface between the tracks (I, Figure 1) the overlap did not temper the martensite in C condition (Figure 1b) because carbide is the main constituent (Figure 1d).

Figure 2 presents the X-ray diffractogram of the surface treated using condition C. The spectrum of condition B was suppressed in Figure 2 because if presents only $\alpha$ (ferrite) phase. Each peak was identified according to the interplanar spacing (d) and the position (2-theta), however the effect of multiple reflections and the fluorescence effect of the copper radiation in iron render difficult the indexation. A more refined view of the interest interval is presented at up-right corner of Figure 2, showing the peaks associated to the Miller index of ferrite ($\alpha$) and cementite ($Fe_3C$) phases. The X-ray diffraction analyses corroborated the assumption of cementite after LSH in condition C, although the volumetric fractions could not be estimated because of the crystallographic texture.

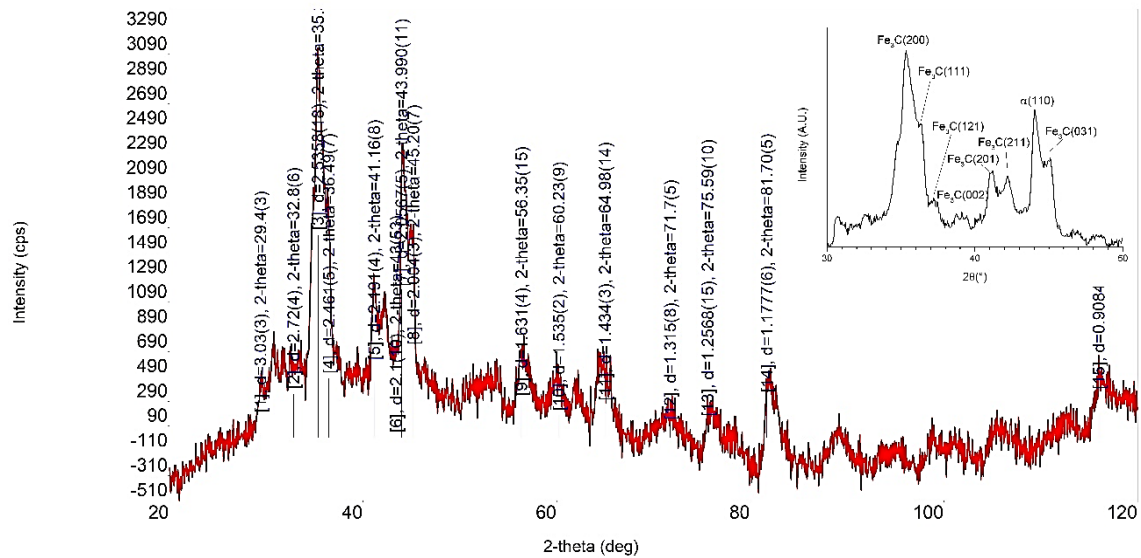

Figure 2. X-ray diffractogram of the surface C. The detail at up right corner is a zoom around the main peaks identifying both cementite and ferrite peaks.

**Hardness profile**

Figure 3 presents the Vickers hardness profile as a function of the depth (z) for B and C conditions. In the figure, the approximate *loci* of each region, HR, PTR and BM, are indicated together with the range of the base material hardness range ($HV_{BM}$).

Two of indentations in the HR of condition B and one near to the interface (I) resulted in hardness values well below than the value obtained at z = 0.2 mm, in the middle of PTR (Figure

3). The absence of a protective gas shielding created a carbon depletion in the HR during remelting, as already seen after laser processing of steels [6, 11]. The maximum hardness of 280 HV is then 0.2 mm from the surface. The values observed in Condition B in BM are slightly above the $HV_{BM}$ as a consequence of some collateral heating of the substrate [3]. Collateral heating also appears to be responsible for the small difference between the hardnesses at 0.58 mm below the surface of parts B and C (Figure 3). Piece B may have heated a little more than part C due to the transformation of cementite, which explains hardness B being slightly above BM and vice versa for C.

Figure 3 also presents the HV profile for C condition where the highest value was obtained 82 µm below the surface (330 HV). The carburization of the exposed surface improved the hardenability of the sample up to approximately the interface (I, Figure 1d). As proposed before, the enthalpy necessary to obtain cementite reduce the total energy for a depth hardenability. Consequently, the HV profile from PTR to BM in Condition C is below the trend line of Condition B.

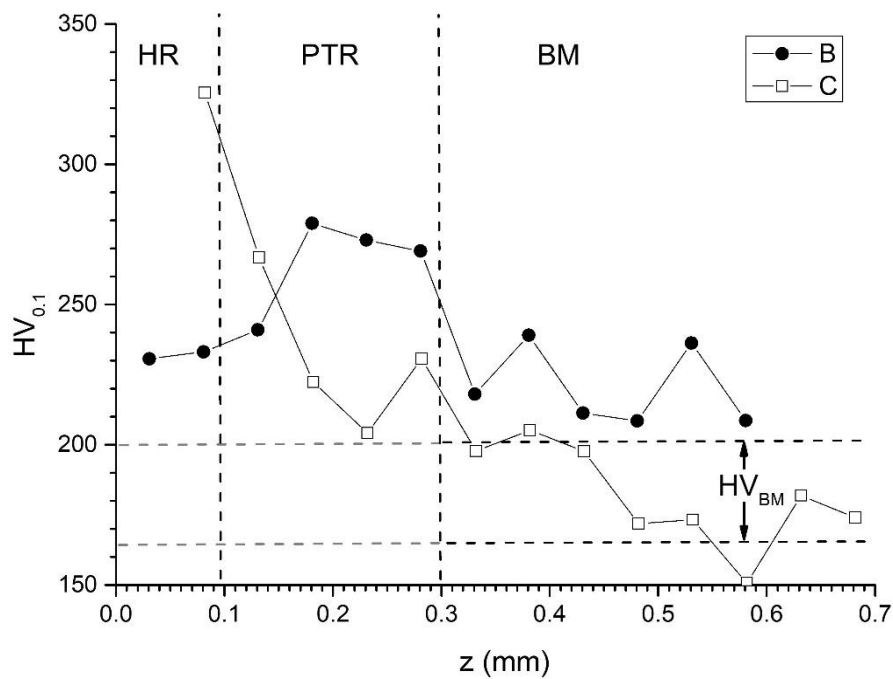

Figure 3. Vickers Hardness profile as a function of depth (z) for conditions B and C. The acronyms are cited in the text.

**Wear behavior**

Reciprocating abrasive wear allowed measuring the resistance to motion called coefficient of friction (COF) as given in Figure 4. The curves were representative of three tests for each experimental condition. Five curves are superposed for comparison: S (substrate, untreated), B (bare, LSH without C) and C (carbon layer LSH), these later with a reciprocating movement transversal (T) or longitudinal (L) to the laser tracks. After an initial period of about 500 s, the curves S, B(T) and B(L) presents some similarities. The S-curve, however, presented a wide variation of COF during the tests. According to Milan et al. [12] this aperiodic variation of COF during abrasive wear is resulting from a complex interaction of the hard metal ball and the AISI 1020 substrate (S) where temperature changes create oxides and heat treats the base material. In all studied cases, the untreated substrate (S) is the unique where some shaking is apperceived on the equipment case. This variation could also be due to the vibration itself as the sensors are quite sensitive.

In the bare (B) conditions, the COF attained a maximum of 0.55 after 160 s (L) and 0.59 after 660 s (T) (Figure 4). Together with S condition, both B(T) and B(L) presented the highest level of final COF, i.e. $COF_{final}$=0.58±0.02. However, differently from S-tests, the equipment did not present vibration. For a while, up to 420 s, the tests transversal to the laser tracks B(T) shown a COF less than B(T), however, after this, the B(T) surpassed any value of B(L). A possible explanation comes from the topography of the treated surfaces. When the tests starts the hard metal ball (counterpart) bumped on the top of laser-fabricated mountains. Then, after a given time, these tips were broken and entered as a third body in the tribological system. On the other hand, B(L) laser tracks were aligned to the ball motion. After a given time, the tracks were less rough and the debris are free to run though the laser tracks (valeys). The hardness of the HR zone (Figure 3) in both B(L) and B(T) are not too above the substrate values and the COF procedure are similar to the untreated surface (S).

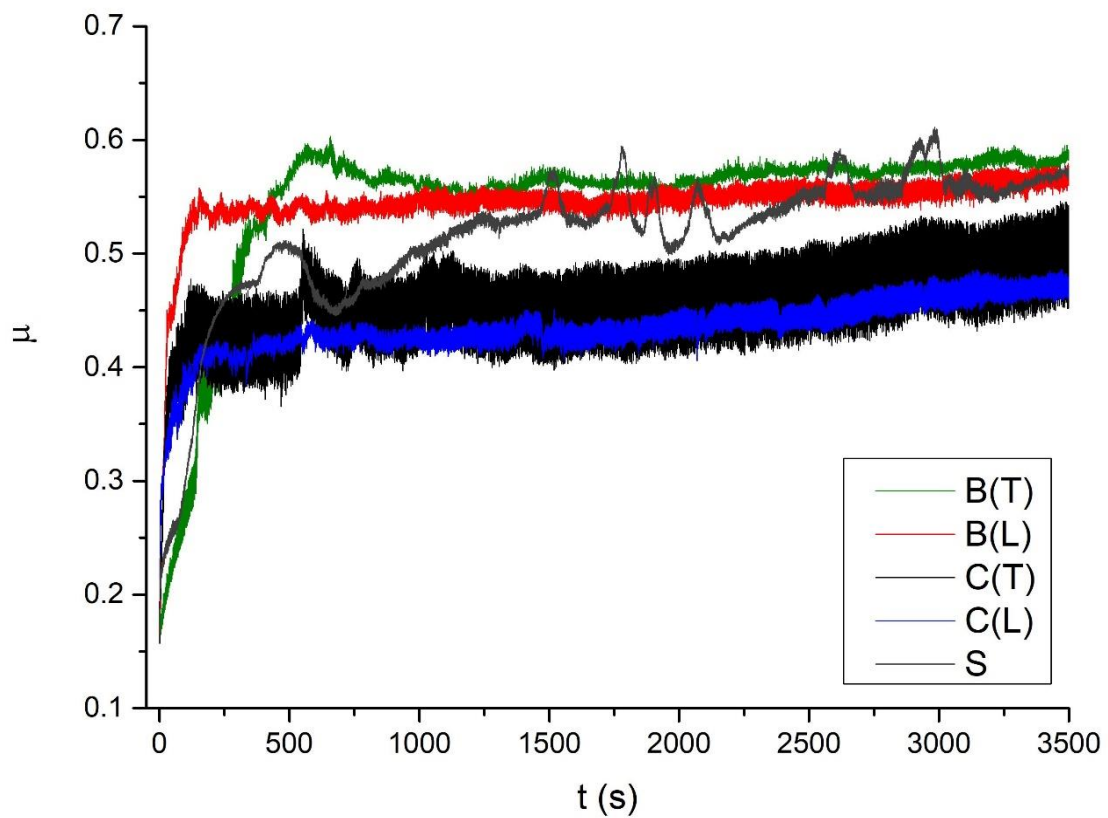

Figure 4. Coefficient of friction as a function of time for the conditions S (substrate, untreated), B and C. The letters L and T mean longitudinal and transversal to the laser tracks.

Figure 4 also presents the evolution of COF as function of time when the surface is pre-coated with carbon before LSH: C(T) and C(L). The C-curves in the figure are quite similar which indicates COF is about the same disregarding if the reciprocating motion is transversal (T) or longitudinal (L) to the laser tracks. The wide dispersion of COF in C(T) condition compared to C(L) indicates that the bumping at the mountain tops created some noise. Although this noise is unperceived during the tests, no shaking the equipment. At the end of the tests, the final

COF for C-conditions is approximately $COF_{final}=0.45\pm0.05$. The coefficient of friction is not a material property but a value liked to many boundaries conditions, however the COFs obtained here are similar to those reported by Dudley D. Fuller (Columbia University) data boards [13].

Figure 5 presents some images representative of the surface damage after the reciprocating tests. The untreated surface (S) presents a scratch after 3500 s with a width of approximately 0.26 mm. As measured by the equipment profilemeter, the resulting wear was $2.81\times10^{-5}$ $mm^3$/N/m for the S-condition. In the other conditions, the equipment was unable to estimate the final wear because of the surface waviness and roughness. The scratch marks are very irregular in both B and C conditions, unrespect to the direction L or T, and some images (as these in Figure 5) indicate that some of hard metal was welded to the scratched surfaces.

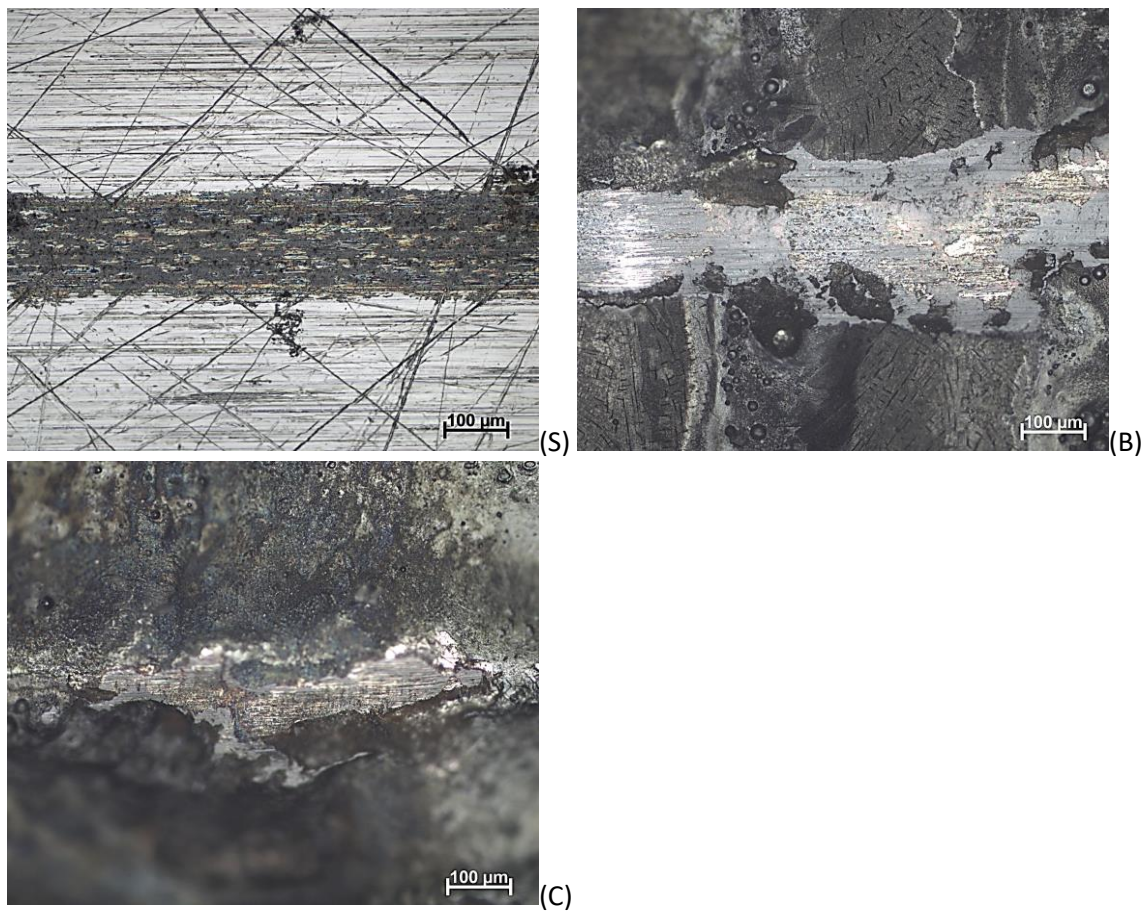

Figure 5. Representative images of the surface damage for the conditions S (substrate, untreated), B and C.

When comparing the surface wear in the hard metal spheres after the tests in different conditions (Figure 6) it is clear the LSH steel, even softer than the ball, grinded part of the counter body. The ball damage increased almost linearly from S to C(T) which corroborates the assumptions of cementite third body as part of the tribological system. Cementite hardness is known as being below tungsten carbide hardness (typical constituent of the hard metal ball), i.e. approximately 1000 HV compared to 1300 HV [14]. However, the carburized surface (Figure 5C) takes advantage of the wavy topography of the laser beam fabricated tracks to better grind the counter body, as can be seen in Figure 6.

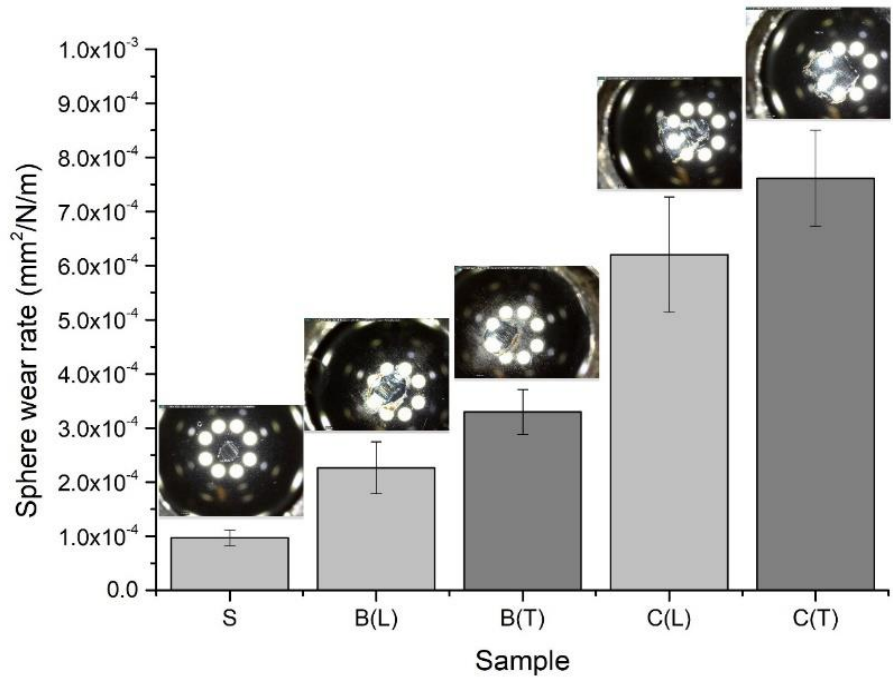

Figure 6. Sphere wear rate as a function of the sample type and relative orientation to the laser path (L and T). The images represent the observed damage after the tribological tests.

## Conclusions

The following conclusions could be drawn:

Laser surface hardening (LSH) is an effective way to increase wear properties of the AISI1020 steel.

In the current experiments, where a fiber laser was used with 150 W beam power, 10 mm/s scanning speed, 2 mm spot diameter and a lateral step of 0.5 mm, carbon coated (C) and bare (B) were produced.

The hardened region (HR) are composed by martensite (B) or by a mixture of ferrite and cementite (C) as determined by light optical microscopy and X-ray diffractometry.

The highest values of Vickers hardness were obtained for carbon coated (C) samples.

The lowest values of the coefficient of friction (COF) were observed in C-condition both when the reciprocating motion is transversal (T) or longitudinal (L) to the laser tracks. The bare (B) condition behave similar to the untreated substrate material (S).

The wear of the LSH surfaces is unperceived, however the hard metal ball (counter body in reciprocating tests) is quite damaged, in particular in the C-condition.

## Acknowledgements

This study was financed in part by the Coordenação de Aperfeiçoamento de Pessoal de Nível Superior - Brasil (CAPES) - Finance Code 001 and the scientific Initiation program of Federal

University of Espirito Santo (Iniciação Científica – Piic/UFES). Thanks are also due to São Paulo Research Foundation (FAPESP) for the grants #2019/26081-3 and #2016/11309-0.

**Author contributions**

MRF: collected the data, performed the analysis

SMC: Conceived and designed the analysis, contributed data or analysis tools, wrote the paper

MSFL: contributed data or analysis tools, wrote the paper

**Conflicts of Interest:** The authors declare no conflicts of interest.

**References**

1 Dinesh Babu, P., Balasubramanian, K. R., Buvanashekaran, G. (2011). Laser surface hardening: a review. International Journal of Surface Science and Engineering, 5(2-3), 131-151.

2 Muthukumaran, G., & Babu, P. D. (2021). Laser transformation hardening of various steel grades using different laser types. Journal of the Brazilian Society of Mechanical Sciences and Engineering, 43(2), 1-29.

3 So, S., & Ki, H. (2013). Effect of specimen thickness on heat treatability in laser transformation hardening. International Journal of Heat and Mass Transfer, 61, 266-276.

4 Ki, H., & So, S. (2012). Process map for laser heat treatment of carbon steels. Optics & Laser Technology, 44(7), 2106-2114.

5 Goia, F., & de Lima, M. (2012). Surface hardening of an AISI D6 cold work steel using a fiber laser. In 18th International Federation for Heat Treatment and Surface Engineering. ASTM International.

6 Maharjan, N., Zhou, W., Zhou, Y., & Wu, N. (2018). Decarburization during laser surface processing of steel. Applied Physics A, 124(10), 1-9.

7 Maharjan, N., Zhou, W., & Wu, N. (2020). Direct laser hardening of AISI 1020 steel under controlled gas atmosphere. Surface and Coatings Technology, 385, 125399.

8 Vaitkus, A., Merkys, A. & Gražulis, S. (2021). Validation of the Crystallography Open Database using the Crystallographic Information Framework. Journal of Applied Crystallography, 54(2), 661-672. doi: 10.1107/S1600576720016532 (BibTeX, plain text)

9 Moradi, M., Moghadam, M. K., & Kazazi, M. (2019). Improved laser surface hardening of AISI 4130 low alloy steel with electrophoretically deposited carbon coating. Optik, 178, 614-622.

10 Hallstedt, B., Djurovic, D., von Appen, J., Dronskowski, R., Dick, A., Körmann, F., ... & Neugebauer, J. (2010). Thermodynamic properties of cementite (Fe3C). Calphad, 34(1), 129-133.

11 Sergeev, N. N., Minaev, I. V., Gvozdev, A. E., Cheglov, A. E., Tikhonova, I. V., Gubanov, O. M., ... & Breki, A. D. (2018). Decarburization and the influence of laser cutting on steel structure. Steel in Translation, 48(5), 313-319.

12 Milan, J. C. G., Carvalho, M. A., Xavier, R. R., Franco, S. D., & De Mello, J. D. B. (2005). Effect of temperature, normal load and pre-oxidation on the sliding wear of multi-component ferrous alloys. Wear, 259(1-6), 412-423.

13 Fuller, D.D., Coefficients of Friction, Available at https://web.mit.edu/8.13/8.13c/references-fall/aip/aip-handbook-section2d.pdf (Accessed Oct.10, 2021)

14 CES Edupack software, Granta Design Limited, Cambridge, UK, 2009.